\documentclass[aps,prb,floatfix,twocolumn,showpacs]{revtex4}
\usepackage{graphics}
\usepackage{epsfig}

\newcommand{\ba}{{\bf a}}

\newcommand{\beq}{\begin{eqnarray}}
\newcommand{\eeq}{\end{eqnarray}}
\newcommand{\beqn}{\begin{eqnarray}}
\newcommand{\eeqn}{\end{eqnarray}}

\begin{document}


\title{Pair Superfluid and Supersolid of Correlated Hard-Core Bosons on a Triangular Lattice}

\author{Hong-Chen Jiang} \affiliation{Kavli Institute for Theoretical Physics, University
of California, Santa Barbara, CA 93106, USA}%
\affiliation{Center for Quantum Information, IIIS, Tsinghua
University, Beijing, 100084, China}

\author{Liang Fu} \affiliation{Department of Physics, Massachusetts Institute of
Technology, Cambridge, MA 02139, USA}

\author{Cenke Xu} \affiliation{Department of Physics, University of California,
Santa Barbara, CA 93106, USA}

\date{\today}

\begin{abstract}

We have systematically studied the hard-core Bose-Hubbard model
with correlated hopping on a triangular lattice using
density-matrix renormalization group method. A rich ground state
phase diagram is determined. In this phase diagram there is a
supersolid phase and a pair superfluid phase due to the interplay
between the ordinary frustrated boson hopping and an unusual
correlated hopping. In particular, we find that the quantum phase
transition between the supersolid phase and the pair superfluid
phase is continuous.

\end{abstract}


\maketitle

\section{Introduction}

The Bose-Hubbard model, as the first explicit example of quantum
phase transition, has been studied extensively with many
techniques. The simplest version of Bose-Hubbard model only
involves a Mott insulator phase and a superfluid phase
\cite{bosehubbard}, and the quantum phase transition between these
two phases can be well-described by a semiclassical
Landau-Ginzburg theory. For example, in two dimension, this
quantum phase transition is either an ordinary 3d XY transition or
a $z = 2$ mean field transition depending on the chemical
potential. In the last few years, it was proposed that various
extended Bose-Hubbard models can have much richer and more exotic
behaviors. For example, a $Z_2$ topological liquid phase has been
discovered in an extended Bose-Hubbard model on the Kagom\'{e}
lattice \cite{balentsz2,kimz2,melkoz2,motrunich}, and in the same
model an exotic quantum phase transition between the $Z_2$ liquid
and a conventional superfluid phase was identified \cite{melko}.
Also, with an extra ring exchange term, it was demonstrated both
numerically and analytically that the Bose-Hubbard model can have
an exotic fractionalized Bose metal phase
\cite{bosemetal1,bosemetal2,bosemetal3}.

With a hard core constraint, $i.e.$ doubly occupied sites are
removed from the Hilbert space, the Bose-Hubbard model is
equivalent to a spin-1/2 model. Due to the rapid development of
numerical techniques, exotic phases have been identified in many
quantum spin-1/2 models as well. For example, based on the
density-matrix renormalization group (DMRG) method, a fully gapped
topological liquid phase has been discovered in the Kagom\'{e}
lattice spin-1/2 Heisenberg model \cite{Jiang2008,white}, as well
as the $J_1 - J_2$ Heisenberg model on the square lattice
\cite{hongchen}.

In this paper, using the DMRG method, we demonstrate that in one
simple extended hard core Bose-Hubbard model, there are three
different interesting phenomena: first of all, there is a
supersolid (SS) phase, where there is a coexistence of the
off-diagonal long range order of boson creation operator, and a
boson density wave order. Secondly, there is a pair superfluid
(PSF) phase, where $\langle b_i \rangle = 0$ while $\langle b_i
b_{i+\alpha} \rangle \neq 0$. This pair superfluid phase is an
analogue of the charge-$4e$ superconductor that was discussed
lately \cite{berg4e,moon4e}. Thirdly, we show that there is a
continuous quantum phase transition between the supersolid and the
pair superfluid phase.

\section{Model Hamiltonian}

\begin{figure}[tbp]
\centerline{
\includegraphics[height=1.0in,width=2.8in]{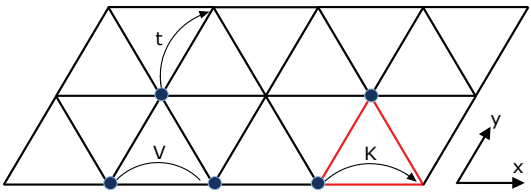}
} \caption{(Color online) Illustration of the 3-leg ladder
geometry with unit vectors $\hat{x}=(1,0)$ and
$\hat{y}=(\frac{1}{2},\frac{\sqrt{3}}{2})$. The boson hopping
strength is $t$, and the repulsive interaction is $V$. The
correlated hopping term $K$ moves a hard-core boson from one site
to another one, depending on the number of boson in the third site
in the same small triangle.} \label{Fig:TriangularLattice}
\end{figure}

We consider a hard-core Bose-Hubbard model with a correlated
hopping on a triangular lattice
\begin{eqnarray}
H &=&t\sum_{\langle ij\rangle} \left(b^+_i b_j + h.c.\right) +
V\sum_{\langle ij\rangle}n_i n_j\nonumber\\
&-& K\sum_{ijk\in\bigtriangleup} \left(n_i b^+_j b_k +
h.c.\right),\label{Eq:ModelHamiltonian}
\end{eqnarray}
where $b^+_i$ ($b_i$) is the boson creation (annihilation) operator
and $n_i$ is the boson number operator on site $i$. In this
Hamiltonian $t$ is the ordinary nearest-neighbor (NN) boson hopping
amplitude, and $V$ is the NN repulsive interaction. $K$ is a
correlated hopping term, and $ijk\in\triangle$ are three sites in a
small triangle (shown in Fig.\ref{Fig:TriangularLattice}) of the
lattice. We will see that the presence of the $K$ term significantly
enriches the phase diagram of this model. In the numerical
simulations, for simplicity, we will always set $t=1$ as the unit of
energy, and focus on the case with $V>0$ and $K>0$. Notice that here
the boson hopping term is frustrated.

We determine the ground-state phase diagram of the model Hamiltonian
Eq.~(\ref{Eq:ModelHamiltonian}) by extensive and highly accurate
DMRG\cite{White1992,Stoudenmire2011} simulations. In particular, we
consider a system with total number of sites $N=L_x\times L_y$,
which are spanned by multiples $L_x \hat{x}$ and $L_y \hat{y}$ of
the unit vectors $\hat{x} = (1,0)$ and $\hat{y} = (\frac{1}{2},
{\frac{\sqrt{3}}{2}})$. For the DMRG calculation, we consider both a
cylinder boundary condition (CBC) and a fully periodic boundary
condition. Here, CBC means open boundary condition along $L_x$
direction, while periodic boundary condition along $L_y$ direction.
This allows us to work on numerous cylinders with much larger system
size to reduce the finite-size effect for a more reliable
extrapolation to the thermodynamic limit. We keep more than $m=6000$
states in each DMRG block for most systems, which is found to give
excellent convergence with tiny truncation errors that can be fully
neglected.

\begin{figure}[tbp]
\centerline{
    \includegraphics[height=2.4in,width=3.4in]{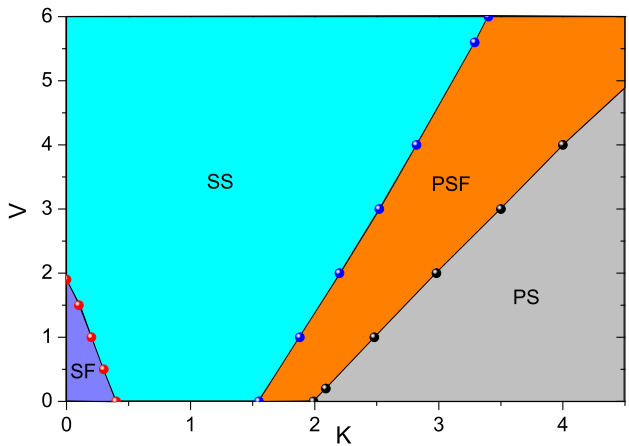}
    }
\caption{(Color online) Ground state phase diagram for the
correlated hard-core Bose-Hubbard model (Eq.
(\ref{Eq:ModelHamiltonian})) in a triangular lattice at filling
$\rho=1/6$, as determined by accurate DMRG calculations on
cylinders with $L_y$ up to $9$. Changing the coupling parameters
$V$ and $K$, four different phases are found, including the
superfluid (SF), supersolid (SS), pair superfluid (PSF), and the
phase separation (PS).} \label{Fig:F16PhaseDiagram}
\end{figure}

\section{Phase Diagram}

The main result of this paper is illustrated in the phase diagram of
the model (Eq.(\ref{Eq:ModelHamiltonian})) at filling
$\rho=\frac{1}{6}$, as shown in Fig.\ref{Fig:F16PhaseDiagram},
obtained by extensive DMRG studies on numerous cylinders with
$L_y=3-9$. The nature of the ground state of the model Hamiltonian
Eq.(\ref{Eq:ModelHamiltonian}) at half filling has already been
studied previously without considering the correlated hopping term
$K$, where a supersolid phase was found to be stable over a wide
range of interaction strength\cite{ashvinSS1,Jiang2009,Wang2009}. In
the SS phase there is a long range order of both the boson creation
operator and the boson density wave. Further study\cite{Jiang2011}
shows that this SS phase will also survive at low boson density,
such as $\rho=\frac{1}{6}$, when the repulsive interaction is strong
enough. For weak interaction, this SS phase will give way to the
simple atomic superfluid phase.

In our current paper, we demonstrate that the unfrustrated
correlated hopping term will compete with the repulsive
interaction and lead to an interesting phase diagram. In
particular, with intermediate strength of $K$, the SS phase is
driven into a uniform pair superfluid phase where $\langle b_i
\rangle = 0$, while $\langle b_i b_{i+\alpha} \rangle \neq 0$.

\begin{figure}[tbp]
\centerline{
    \includegraphics[height=4.8in,width=3.4in]{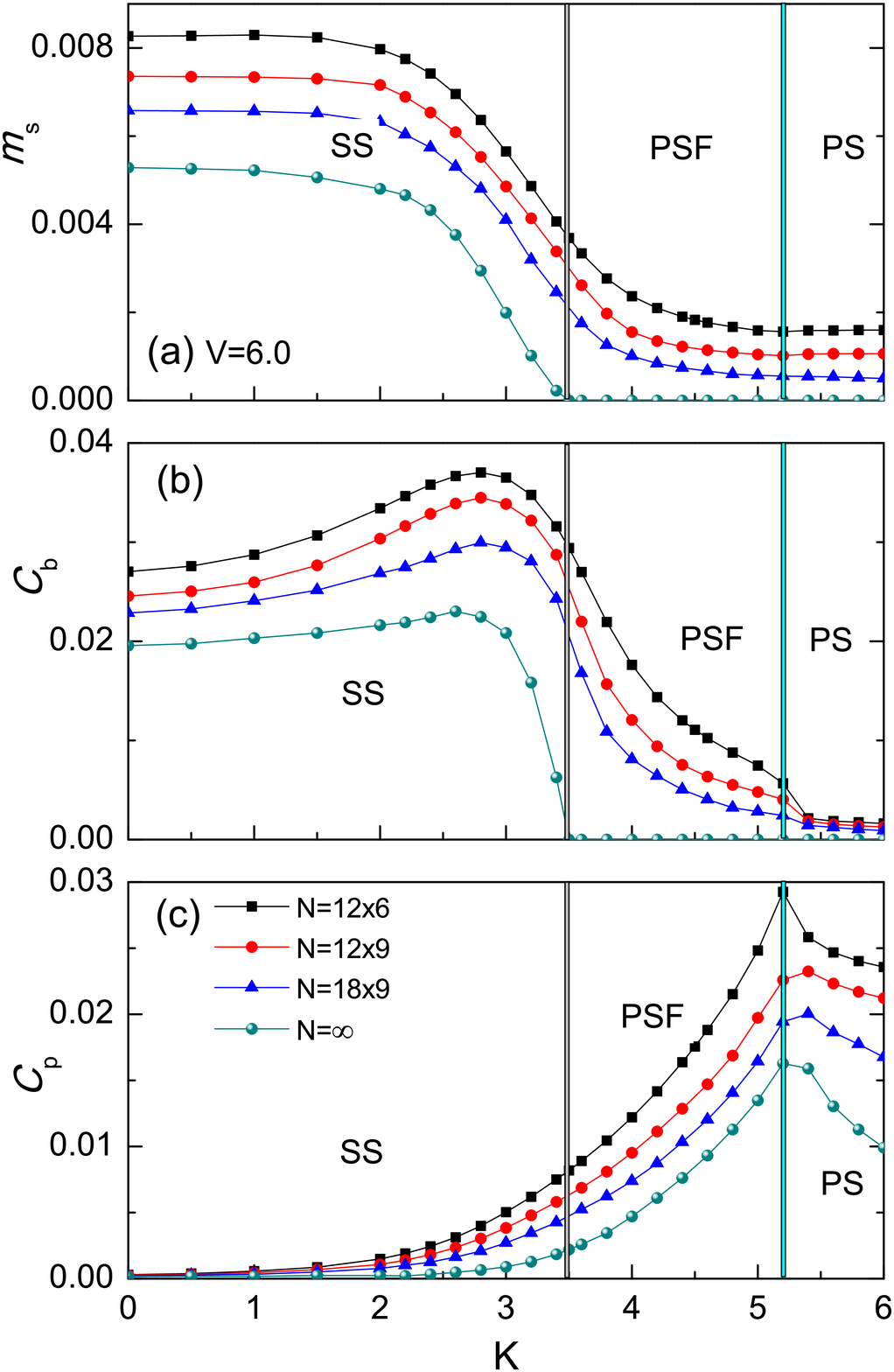}
    }
\caption{(color online) (a) The density structure factor order
parameter $m_s$, (b) atomic condensate density $C_b$ at momentum
$k_1=(4\pi/3,0)$, and (c) pair condensate density $C_p$ at
momentum $k_0=(0,0)$, as functions of $K$ at $V=6.0$ and filling
$\rho=1/6$ , respectively, with system size $N=12\times 6$,
$12\times 9$, $18\times 9$, and the corresponding extrapolations
in the thermodynamic limit.} \label{Fig:V6SkHkF16}
\end{figure}

\begin{figure}[t]
\centerline{
    \includegraphics[height=4.8in,width=3.4in]{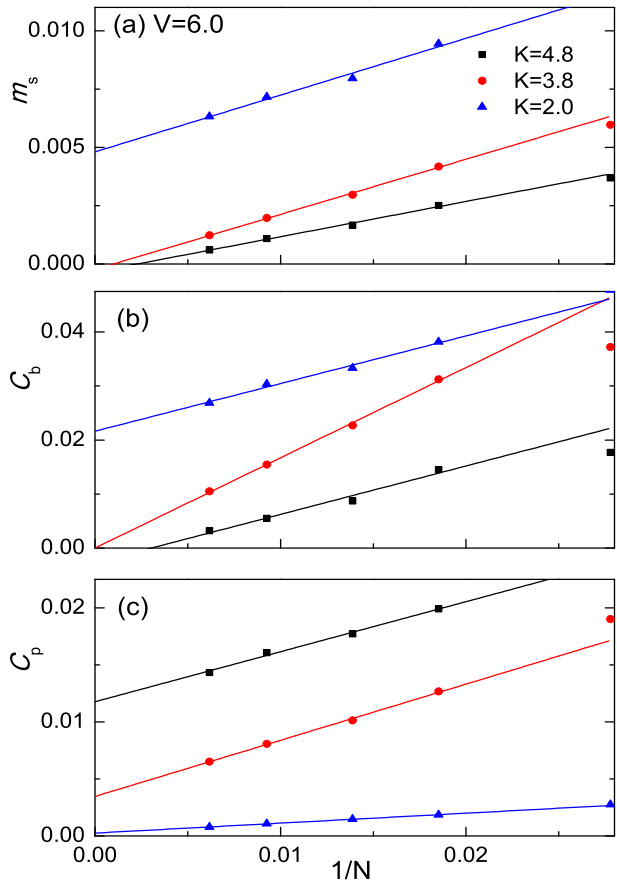}
    }
\caption{(color online) Examples of finite-size scaling of (a) the
density structure factor order parameter $m_s$ and (b) atomic
condensate density $C_b$ at momentum $k_1=(4\pi/3,0)$, as well as
(c) the pair condensate density $C_p$ at momentum $k_0=(0,0)$, for
different $K$ at $V=6.0$ and filling $\rho=1/6$, with system size
up to $18\times 9$.} \label{Fig:V6SkHkF16Scaling}
\end{figure}

To analyze the ground state properties of the system, we calculate
both the density structure factor
\begin{eqnarray}
S(\textbf{k})=\frac{1}{N}\sum_{ij}e^{i\textbf{k}\cdot(\textbf{r}_i-\textbf{r}_j)}\langle
(n_i-\rho)(n_j-\rho)\rangle,
\end{eqnarray}
and the momentum distribution function%
\begin{eqnarray}
M_b(\textbf{k})=\frac{1}{N}\sum_{ij}e^{i\textbf{k}\cdot(\textbf{r}_i-\textbf{r}_j)}\langle
b^+_i b_j\rangle,
\end{eqnarray}
where $\rho$ is the filling factor of the system. We also
calculate the pair superfluid structure factor%
\begin{eqnarray}
M^\alpha_p(\textbf{k})=\frac{1}{N}\sum_{ij}e^{i\textbf{k}
\cdot(\textbf{r}_i-\textbf{r}_j)}\langle \Delta^{+\alpha}_i
\Delta^\alpha_j\rangle,
\end{eqnarray}
to characterize the pair superfluid phase. Here
$\Delta^\alpha_i=b_ib_{i+\alpha}$ is the nearest-neighbor pair
annihilation operator along $\alpha$ direction, with
$\alpha=\hat{x}$, $\hat{y}$ or $\hat{y}-\hat{x}$.

In both the SF phase and supersolid phase, the obtained $S(k)$ and
$M_b(k)$ show Bragg peaks at the corners of the hexagonal
Brillouin zone, e.g., at $k_1 = (\pm 4\pi/3,0)$. In particular, at
small $K$, both the peak of the density structure factor $S(k)$
and momentum distribution function $M(k)$ are very sharp (not
shown). As shown in Fig.\ref{Fig:V6SkHkF16}, at large $V=6.0$,
with the increase of $K$, $S(k_1)$ decreases continuously, while
$M_b(k_1)$ increases at first and then decreases with larger $K$.
Eventually both $S(k_1)$ and $M_b(k_1)$ becomes very weak beyond
certain critical value of $K$. On the other hand, $M^\alpha_p(k)$
shows peaks at zero momentum, i.e., $k_0=(0,0)$, and will increase
with $K$, $M^\alpha_p(k)$ increases monotonically before phase
separation. One can obtain the corresponding order parameters in
the thermodynamic limit based on the finite-size scaling of the
peak values $S(k_1)$, $M_b(k_1)$ and $M^\alpha_p(k_0)$.
Specifically, the structure factor order parameter and the boson
condensate density can be determined by $m_s=S(k_1)/N$ and $
C_b=M_b(k_1)/N$, while the total pair condensate density is given
by $C_p=\sum_{\alpha} C^\alpha_p=\sum_{\alpha}M^\alpha_p(k_0)/N$.
Examples of these order parameters are shown in
Fig.\ref{Fig:V6SkHkF16} as a function of correlated hopping $K$
with $N=12\times 6$, $12\times 9$ and $18\times 9$.

\begin{figure}[tbp]
\centerline{
    \includegraphics[height=4.8in,width=3.4in]{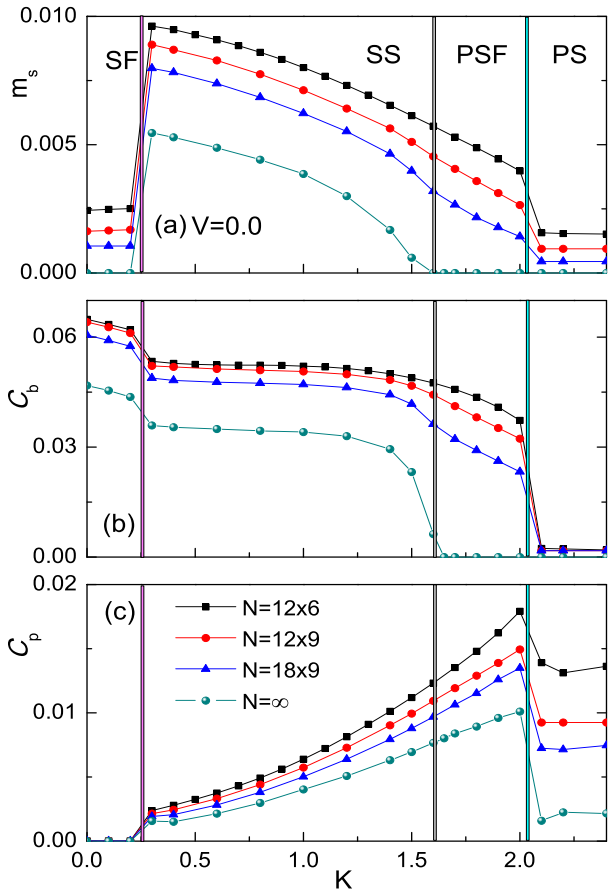}
    }
\caption{(color online) (a) The density structure factor order
parameter $m_s$, (b) atomic condensate density $C_b$ at momentum
$k_1=(4\pi/3,0)$, and (c) the pair condensate density $C_p$ at
momentum $k_0=(0,0)$, as functions of $K$ at $V=0.0$ and filling
$\rho=1/6$, respectively, with system size $N=12\times 6$,
$12\times 9$ and $18\times 9$, as well as the corresponding
extrapolations in the thermodynamic limit.} \label{Fig:V0SkHkF16}
\end{figure}

Nonzero $m_s$ and $C_b$ in the thermodynamic limit correspond to
the diagonal long-range order (LRO) and off-diagonal long-range
order (ODLRO), respectively. Examples of the finite-size scaling
are shown in Fig.\ref{Fig:V6SkHkF16Scaling} by plotting $m_s$,
$C_b$ and $C_p$ as functions of $1/N$, at $V=6.0$ and different
$K$, using quasi-2D lattice geometry with system size up to
$N=18\times 9$. The obtained order parameters extrapolated to the
thermodynamic limit are presented in Fig.\ref{Fig:V6SkHkF16} (dark
cyan sphere) and Fig.\ref{Fig:V6SkHkF16Scaling}. It is worth
noting that after extrapolation both $m_s$ and $C_b$ continuously
decrease to zero simultaneously at the critical $K$, and the
system enters the pair superfluid phase in the thermodynamic
limit, indicating a continuous phase transition. However, $C_p$ is
nonzero in both the SS and PSF, and it always increases
monotonically with $K$ (See Fig. \ref{Fig:V6SkHkF16}(c)) before
phase separation. Specifically, the finite-size scaling for $C_p$
at $K=2.0$ in the supersolid phase gives us a small but finite
value $C_p\approx 2.5\times 10^{-4}$ (around $2\%$ of $C_p$ at
$K=4.8$), as shown in Fig.\ref{Fig:V6SkHkF16Scaling}(c).

With the decrease of the repulsive interaction $V$, both the
supersolid phase and pair superfluid phase become weaker and move
to smaller $K$. In particular, a superfluid phase appears at small
$K$ locating at the left bottom corner in the phase diagram (see
Fig.\ref{Fig:F16PhaseDiagram}). Compared with the supersolid and
pair superfluid phase at large $V$, the pair condensate density at
$k_0=(0,0)$ becomes zero, i.e., $C_p(k_0)=0$. Even with the
absence of the repulsive interaction $V$, the atomic superfluid
phase is completely suppressed with the increase of $K$ and give
way to the supersolid phase through a first-order transition
around $K=0.3$. This can be seen clearly in
Fig.\ref{Fig:V0SkHkF16}, in which $m_s$ and $C_p$ jump from zero
to a finite value in the thermodynamic limit, while $C_b$
encounters a sharp drop to a smaller but still finite value. It is
interesting to note that the pair superfluid phase still remains
robust in a finite parameter region $K\approx 1.6-2.0$, even in
the absence of the repulsive interaction $V$, due to the
competition between $t$ and $K$. Finally, the system becomes phase
separated when $K$ becomes dominant.

\section{Theoretical understanding}

In this section, we develop a simple Landau-Ginzburg theory to describe the
supersolid, superfluid and paired superfluid phases on the
triangle lattice, including the phase transitions between them.
The dispersion of a boson with frustrated hopping in triangular
lattice has two minimums at $\vec{K}$ and $-\vec{K}$, where
$\vec{K}=(4\pi/3, 0)$. Therefore we introduce the two $U(1)$ order
parameters $\phi_{1,2}$ for boson condensate at $\pm \vec{K}$
\begin{eqnarray}
b^\dagger (\vec{x}) \sim \phi_1(\vec{x}) e^{ i \vec{K} \cdot
\vec{x}} + \phi_2(\vec{x}) e^{ - i \vec{K} \cdot \vec{x}}
\label{phi1phi2}
\end{eqnarray}
Under time reversal and lattice symmetries, $\phi_{1,2}$ transform as follows:
\begin{eqnarray}
&T_i&: \; \phi_1 \rightarrow \phi_1 e^{- i 2 \pi/3}, \;
\phi_2 \rightarrow \phi_2 e^{i 2 \pi/3}, \; i=1,2,3 \nonumber \\
&C_6&: \; \phi_1 \rightarrow \phi_2, \; \phi_2 \rightarrow \phi_1 \nonumber \\
&M_x&: \; \phi_1 \rightarrow \phi_2, \phi_2 \rightarrow \phi_1 \nonumber \\
&\Theta &: \;  \phi_1 \rightarrow \phi_2, \; \phi_2 \rightarrow \phi_1 \nonumber \\
&U(1)&: \;  \phi_1 \rightarrow \phi_1 e^{ i \theta}, \;  \phi_2 \rightarrow \phi_2 e^{i \theta},
\end{eqnarray}
Here $T_i$ are translations by the three unit vector $\ba_i$ of
the triangle lattice: $\ba_1 = (1, 0), \; \ba_2=(-1/2,
\sqrt{3}/2), \; \ba_3= (-1/2, -\sqrt{3}/2)$. $C_6$ is $\pi/3$
rotation. $M_{x}$ is reflection about $x$ axis. $\Theta$ is
time-reversal transformation. $U(1)$ is the global phase
transformation.

The supersolid phase is characterized by {\it two} nonzero condensate order parameters
$\phi_1$ and $\phi_2$, which have momentum $\vec K$ and $- \vec K$ respectively.
The coexistence of these two superfluid order
 necessarily induces a nonzero density wave order $\rho \sim \phi_1^* \phi_2$
 at momentum $2 \vec K = - \vec K$, where we have
used the fact $3 \vec K = 0$ up to a reciprocal lattice vector in
triangular lattice. The presence of both superfluid and
density-wave order taken together signals a supersolid phase, as we found
numerically. On the other hand, the paired superfluid phase is characterized by
a nonzero pair condensate order parameter $\Delta$ at zero momentum, whereas
both $\phi_1$ and $\phi_2$ are disordered.

Since the supersolid phase has a lower symmetry than the paired
superfluid phase, the phase transition can be understood as the
development of long-range single boson superfluid in the paired
superfluid (disordered) phase within the framework of
Landau-Ginzburg theory in 2+1 dimension. The effective
Landau-Ginzburg Lagrangian, whose form is dictated by the symmetry
property (6), is given by \beq
\mathcal{L} &=&  \frac{r}{2} ( \phi^*_1 \phi_1  + \phi^*_2 \phi_2 )  - (\Delta \phi_1^* \phi_{2}^*  +  \Delta^* \phi_1 \phi_{2}  ) \nonumber \\
&+& u  ( |\phi_1|^4  +  |\phi_2|^4 ) + 2u_{12} |\phi_1|^2 |
\phi_2|^2  + v (\phi_1^{*3} \phi_2^3  + \phi_1^{3} \phi_2^{*3} )
\nonumber \eeq As $r$ decreases, $\phi_1$ and $\phi_2$ become
nonzero, which signals the onset of superfluid order as well as
the associated density wave order $\phi_1^* \phi_2$. Note that the
pair order parameter $\Delta$ is nonzero across the transition.
Without loss of generality, $\Delta$ is chosen to be real and
positive. Due to the trilinear coupling term between $\Delta$,
$\phi_1$ and $\phi_2$, $\mathcal{L}$ is minimized by $ \phi_2^*=
\phi_1 \equiv \phi$, where $\phi$ is {\it complex}. The last term
in $S$ is symmetry allowed for the triangular lattice, and locks
the relative phase between $\phi_1$ and $\phi_2$, and pins the
phase of $\rho = \phi_1 \phi_2^* = \phi^2$ to three distinct
values corresponding to three degenerate density wave patterns in
the supersolid phase. In terms of $\phi$, $\mathcal{L}$ is given
by \beq \mathcal{L} = |\partial_\mu \phi|^2 + r_1 | \phi|^2 + u'
|\phi|^4 + v (\phi^6 + \phi^{*6}). \label{phi6}\eeq Except for the
last term, $\mathcal{L}$ is the standard complex scalar field
theory in 2+1 dimension, and this transition belongs to the 3d XY
universality class with order parameter $\phi$. By power-counting,
the sixth-order phase-locking term is strongly irrelevant at the
critical point, thus this transition is continuous.

Interestingly, the above XY transition differs from conventional
paired to single boson superfluid transition, which lies in the
Ising universality class. The distinction arises from the fact
that the superfluid phase studied here has two coexisting (rather
than one) condensate order parameters, of which the relative phase
is a well-defined physical quantity associated with the density
wave order. Our theory can be directly tested by further numerical
studies on the critical exponent for the physical order parameter
at the quantum critical point. For example, the critical exponent
$\nu$ take the value of the ordinary 3d XY transition: $\nu \sim
0.67$. Also, the density wave order parameter $\rho \sim \phi^2$
is a bilinear of $\phi$ in Eq.~\ref{phi6}, thus our theory
predicts that the order parameter $\rho$ has an anomalous
dimension $\eta \sim 1.49$. These predictions can be verified by
further numerical studies on model Eq.~\ref{Eq:ModelHamiltonian}.

We can also interpret the supersolid to pair superfluid transition
in terms of the topological defects inside the supersolid phase.
For convenience, let us rewrite the fields $\phi_1$ and $\phi_2$
introduced in Eq.~\ref{phi1phi2} as follows: \beqn \phi_1 = \phi \
\psi, \ \ \ \phi_2 = \phi^\ast \ \psi. \label{Cpsi} \eeqn $\phi$
and $\phi^\ast$ are complex fields that carry lattice momentum
$\vec{K}$ and $-\vec{K}$ respectively, while $\psi$ carries the
U(1) symmetry of the original boson operator $b_i$. Thus the boson
density wave order parameter is $\rho \sim \phi_1 \phi^\ast_2 \sim
\phi^2$, and the pair superfluid order parameter is $ b_i
b_{i+\alpha} \sim \phi_1 \phi_2 \sim \psi^2$.

In Eq.~\ref{Cpsi}, because $\phi_1$ and $\phi_2$ are both physical
degrees of freedom, $\phi$ and $\psi$ are defined up to a $Z_2$
gauge ambiguity, $i.e.$ the physics is unchanged under
transformation $\phi \rightarrow -\phi$, $\psi \rightarrow -\psi$.
The supersolid phase corresponds to the case where both $\phi$ and
$\psi$ are condensed, while in the pair superfluid phase only
$\psi$ is condensed. Inside the supersolid phase, the smallest
superfluid vortex, has only $\pi-$vorticity, $i.e.$ it is a bound
state between a $\pi-$vortex of $\psi$ and a $\pi-$vortex of
$\phi$. In other words, both $\psi$ and $\phi$ will change sign
after encircling this vortex, while the physical degree of freedom
is unchanged. Notice that the $\pi-$vortex of $\phi$ is a full
vortex of boson density wave order $\rho$, which is equivalent to
a {\it dislocation} of the density wave pattern.

Starting with the supersolid phase, if we want to drive a transition
into the pair superfluid phase, we need to ``melt" the boson density
wave order by condensing its defects. However, since the pair
superfluid phase also has a half quantum $\pi-$vortex, this
transition cannot be driven by condensing the smallest $\pi-$vortex
discussed in the previous paragraph. Instead, it must be driven by
the condensation of the $2\pi-$vortex of $\phi$, which only melts
the boson density wave, but leaves the superfluid stiffness
unaffected. Since the physical boson density wave order parameter
$\rho \sim \phi^2 $ is a bilinear of $\phi$, this transition is
driven by condensing a ``double" dislocation of $\rho$. Thus this
transition is analogous to the transition between pair-density-wave
and charge-4$e$ superconductor discussed in Ref.~\cite{berg4e}.

\section{Summary and Extension}

We have dstudied a hard-core Bose-Hubbard model with an unusual
correlated hopping on a triangular lattice using density-matrix
renormalization group method. In the phase diagram, we discovered a
supersolid phase and a pair superfluid phase, in addition to the
standard superfluid phase. The supersolid and pair superfluid phases
were discussed separately before in different spin models
\cite{ashvinSS1,Jiang2009, Wang2009,wessel}. However, to our
knowledge it is the first time that both these phases are realized
in one model. We also theoretically establish that the phase
transition between the supersolid phase and the pair superfluid
phase is continuous.

If the phase coherence and superfluid stiffness of the pair
superfluid phase are destroyed, then the system most likely enters
a fully gapped $Z_2$ liquid phase with the same topological order
as the toric code model \cite{Kitaev2003}. Presumably this new
transition can be obtained by turning on some extra terms in the
Hamiltonian. We will leave this to future study.

\section{Acknowledgement}

We would like to thank Matthew Fisher, Steve Kivelson, W. Vincent
Liu, Senthil Todadri, and especially Leon Balents and Fa Wang for
insightful discussions. This work was partially supported by the
KITP NSF grant PHY05-51164 and the NSF MRSEC Program under Award No.
DMR 1121053, and the NBRPC (973 Program) 2011CBA00300
(2011CBA00302). Cenke Xu is supported by the Sloan Foundation. Liang
Fu is supported by start-up funds of MIT.


\end{document}